# The Joint Automated Repository for Various Integrated Simulations (JARVIS) for data-driven materials design

Kamal Choudhary[1,2,3], Kevin F. Garrity[1], Andrew C. E. Reid[1], Brian DeCost[1], Adam J. Biacchi[4], Angela R. Hight Walker[4], Zachary Trautt[1], Jason Hattrick-Simpers[1], A. Gilad Kusne[1], Andrea Centrone[4], Albert Davydov[1], Jie Jiang[5], Ruth Pachter[5], Gowoon Cheon[6], Evan Reed[6], Ankit Agrawal[7], Xiaofeng Qian[8], Vinit Sharma[9,10], Houlong Zhuang[11], Sergei V. Kalinin[12], Bobby G. Sumpter[12], Ghanshyam Pilania[13], Pinar Acar[14], Subhasish Mandal[15], Kristjan Haule[15], David Vanderbilt[15], Karin Rabe[15], Francesca Tavazza[1]

1. Materials Measurement Laboratory, National Institute of Standards and Technology, Gaithersburg, MD, 20899, U.S.A.
2. Theiss Research, La Jolla, CA, 92037, U.S.A.
3. Department of Chemistry and Biochemistry, University of Maryland, College Park, MD, 20742, USA.
4. Physical Measurement Laboratory, National Institute of Standards and Technology, Gaithersburg, MD, 20899, U.S.A.
5. Materials and Manufacturing Directorate, Air Force Research Laboratory, Wright–Patterson Air Force Base, OH 45433, USA.
6. Department of Materials Science and Engineering, Stanford University, Stanford, CA, 94305, U.S.A.
7. Department of Electrical and Computer Engineering, Northwestern University, Evanston, IL 60208, U.S.A.
8. Department of Materials Science and Engineering, Texas A&M University, TX, 77843, U.S.A.
9. Joint Institute for Computational Sciences, University of Tennessee, Knoxville, TN, 37996, USA.
10. National Institute for Computational Sciences, Oak Ridge National Laboratory, TN 37831, U.S.A.
11. School for Engineering of Matter, Transport & Energy, Arizona state university, AZ, 85287, U.S.A.
12. Center for Nanophase Materials Sciences, Oak Ridge National Laboratory, TN 37831, U.S.A.
13. Materials Science and Technology Division, Los Alamos National Lab, Los Alamos, NM, 87545, U.S.A.
14. Department of Mechanical Engineering, Virginia Tech, Blacksburg, VA 24061, U.S.A.
15. Department of Physics and Astronomy, Rutgers University, Piscataway, NJ, 08901, USA

**Abstract**

The Joint Automated Repository for Various Integrated Simulations (JARVIS) is an integrated infrastructure to accelerate materials discovery and design using density functional theory (DFT), classical force-fields (FF), and machine learning (ML) techniques. JARVIS is motivated by the Materials Genome Initiative (MGI) principles of developing open-access databases and tools to reduce the cost and development time of materials discovery, optimization, and deployment. The major features of JARVIS are: JARVIS-DFT, JARVIS-FF, JARVIS-ML, and JARVIS-Tools. To date, JARVIS consists of ≈ 40,000 materials and ≈ 1 million calculated properties in JARVIS-DFT, ≈ 1,500 materials and ≈ 110 force-fields in JARVIS-FF, and ≈ 25 ML models for material-property predictions in JARVIS-ML, all of which are continuously expanding. JARVIS-Tools provides scripts and workflows for running and analyzing various simulations. We compare our computational data to experiments or high-fidelity computational methods wherever applicable to evaluate error/uncertainty in predictions. In addition to the existing workflows, the infrastructure can support a wide variety of other technologically important applications as part of the data-driven materials design paradigm. The JARVIS datasets and tools are publicly available at the website: https://jarvis.nist.gov .

**Corresponding author:** Kamal Choudhary ( kamal.choudhary@nist.gov )

**Introduction**

The Materials Genome Initiative (MGI)[1,2] was introduced in 2011 to accelerate materials discovery using computational[3-9], experimental[10-13] and data analytics[14-16] approaches. The MGI has revolutionized several fields for materials-applications, such as batteries[17], thermoelectrics[18], and alloy-design[19], thorough open-access public database and tool development[20]. The MGI encourages systematic Process-Structure-Property-Performance (PSPP)[21]-based efficient design-approaches rather than Edisonian trial-error methods[22].

Especially in the field of computational materials design, quantum mechanics-based density functional theory (DFT)[23] has proven to be an immensely successful technique, and several databases of automated DFT calculations are widely used in materials design applications. Despite their successes, existing DFT databases face limitations due to issues intrinsic to conventional DFT approaches, e.g. the generalized gradient approximation of Perdew-Burke-Ernzerhof (GGA-PBE)[23,24]. Drawbacks of the existing DFT databases include non-inclusion of van der Waals (vdW) interactions[8], bandgap underestimations[25], non-inclusion of spin-orbit coupling[7], overly simplifying magnetic ordering[26], neglecting defects[27] (point, line, surface and volume), unconverged computational parameters such as k-points[28], ignoring temperature effects[29] (generally DFT calculations are performed at 0 K), lack of layer/thickness-dependent properties of low dimensional materials[30], and lacking interfaces/heterostructures of materials[31], all of which can be critical for realistic material-applications. Additionally, there are several other computational approaches, such as classical force-field (FF)[32], computational microscopy, phase-field (PF), CALculation of PHAse Diagrams (CALPHAD)[33], and Orientation Distribution Functions (ODF)[34] which lack the integrated tools and databases that have been developed for DFT-based computational approaches. Finally, the integration of computational approaches with experiments, the application of statistical uncertainty analysis, and the implementation of data analytics and artificial intelligence (AI) techniques require significant developments to meet the goals set forth by the MGI.

Some of the notable materials databases are: Automatic-FLOW for Materials Discovery (AFLOW)[3], Materials-project[4], Khazana[17], Open Quantum Materials Database (OQMD)[5], Novel Materials Discovery (NOMAD)[9], Computational Materials Repository (CMR)[35], NIMS-MatNavi[36], NREL-MatDB[37], Inorganic Crystal Structure Database (ICSD)[38], Materials-Cloud[39], Citrine[40], OpenKIM[41], Predictive Integrated Structural Materials Science (PRISMS)[42], and Phase-Field hub (PFhub)[43]. Some of the commonly used computational-tools are Python Materials Genomics (PYMATGEN)[44], Atomic Simulation Environment (ASE)[45], Automated Interactive Infrastructure and Database (AIIDA)[6] and MPinterfaces[46]. The data most commonly included in these databases consists of crystal structures, formation energies, bandgaps, elastic constants, Poisson ratios, piezoelectric constants, and dielectric constants. These material properties can be used directly to screen for potentially interesting materials for a given application as candidates for experimental synthesis and characterization, as well as part of a PSPP design approach to better understand the factors driving material performance. Beyond the directly calculated material properties mentioned above, several selection metrics are also being developed to aid materials design, such as scintillation attenuation length[47], thermoelectric complexity factor[48], spectroscopy limited maximum efficiency[49,50], exfoliation energy[8], and spin-orbit spillage[7,26,51]. Akin to DFT-like standard computational approaches that are used as screening tools for experiments, machine learning (ML) [14-16,52] models for materials design are being developed as pre-screening tools for other conventional computational methods such as DFT. In addition, ML tools are proposed to accelerate experimental methods directly based on computational data[53]. All of the above developments show immense promise for accelerating materials design.

The principles mentioned above constitute the foundations of the Joint Automated Repository for Various Integrated Simulations (JARVIS) (https://jarvis.nist.gov) infrastructure, a set of databases and tools to meet some of the current material-design challenges. The main components of JARVIS are: JARVIS-DFT, JARVIS-FF, JARVIS-ML, and JARVIS-Tools. JARVIS is developed and hosted at the National Institute of Standards and Technology (NIST)[54] as part of the MGI. A detailed documentation webpage for the database is available at: https://jarvis-materials-design.github.io/dbdocs/.

Started in 2017, JARVIS-DFT[7,8,25-27,30,31,49,53,55] is a repository based on DFT calculations that mainly uses the vdW-DF-OptB88 van der Waals functional[56]. The database also uses beyond-GGA approaches for a subset of materials, including the Tran-Blaha modified Becke-Johnson (TBmBJ) meta-GGA[57], the hybrid functional PBE0, the hybrid range-separated functional Heyd-Scuseria-Ernzerhof (HSE06), Dynamical Mean Field Theory (DMFT), and $G_0W_0$. In addition to hosting conventional properties such as formation energies, bandgaps, elastic constants, piezoelectric constants, dielectric constants, and magnetic moments, it also contains previously unavailable datasets, such as exfoliation energies for van der Waals bonded materials, the spin-orbit coupling (SOC) spillage, improved meta-GGA bandgaps, frequency-dependent dielectric functions, the spectroscopy limited maximum efficiency (SLME), infrared (IR) intensities, electric field gradients (EFG), heterojunction classifications, and Wannier tight-binding Hamiltonians. These datasets are compared to experimental results wherever possible to evaluate their accuracy as predictive tools. JARVIS-DFT also introduced protocols such as automatic k-point convergence, which can be critical for obtaining precise and accurate results. JARVIS-DFT is distributed through the website: https://jarvis.nist.gov/jarvisdft/ .

The JARVIS-FF[27,58] database, also started in 2017, is a repository of classical force-field/potential computational data intended to help a user select the most appropriate force-field for a specific application. Many classical force-fields are developed for a particular set of properties (such as energies), and may not have been tested for properties not included in training (such as elastic constants, or defect formation energies). JARVIS-FF provides an automatic framework to consistently calculate and compare basic properties, such as the bulk modulus, defect formation energies, phonons, etc., that may be critical for specific molecular-dynamics simulations. JARVIS-FF relies on DFT and experimental data to evaluate accuracy. JARVIS-FF is distributed through the website: https://jarvis.nist.gov/jarvisff/ .

The JARVIS-ML[49,53,55,59,60] is a repository of machine learning (ML) model parameters, descriptors, and ML-related input and target data. JARVIS-ML introduced Classical Force-field Inspired Descriptors (CFID) in 2018 as a universal framework to represent a material's chemistry-structure-charge related data. With the help of CFID and JARVIS-DFT data, several high-accuracy classification and regression ML models were developed, with applications to fast materials-screening and energy-landscape mapping. Some of the trained property models include formation energies, exfoliation energies, bandgaps, magnetic moments, refractive indexes, dielectric constants, thermoelectric performance, and maximum piezoelectric and infrared modes. Also, several ML interpretability analyses have provided physical-insights beyond intuitive materials-science knowledge[59]. These models, the workflow, the datasets, etc. are disseminated to enhance the transparency of the work. Recently, JARVIS-ML was expanded to include ML models to analyze STM-images in order to directly accelerate the interpretation of experimental images.

Graph convolution neural network models are currently being developed for automated handling of images and crystal-structure analysis in materials science. JARVIS-ML is distributed through the website: https://jarvis.nist.gov/jarvisml/ .

JARVIS-Tools is the underlying computational framework used for automation, data-generation, data-handling, analysis and dissemination of all the above repositories. JARVIS-Tools uses cloud-based continuous integration, low-software dependency, auto-documentation, Jupyter and Google-Colab notebook integration, pip installation and related strategies to make the software robust and easy to use. JARVIS-Tools also hosts several examples to enable a user to reproduce the data in the above repositories or to apply the tools for their own applications. JARVIS-Tools are provided through the GitHub page: https://github.com/usnistgov/jarvis.

While JARVIS has some features in common with existing DFT-based computational databases, we note that there are several features currently unique to the JARVIS framework. First, JARVIS has a tight integration between FF and DFT techniques. Second, JARVIS includes CFID ML learning descriptors and several ML models based on those descriptors, including solar-cell efficiency, thermoelectrics, exfoliation energies, infrared active modes, and refractive index etc. Finally, JARVIS-DFT itself features heavy use of a van der Waals functional, a 2D materials database, a STM image database, spin-orbit calculations, spin-orbit spillage, solar cell efficiency, meta-GGA functional calculations, other post-GGA electronic structure calculations, 2D heterostructure design app and a Wannier function database. We also provide REST-API framework for users to download and upload materials data using JARVIS-API.

This paper is organized as follows: 1) we introduce the main computational techniques, organized by the time and length scales, 2) we illustrate JARVIS-Tools and its functionalities, 3) we discuss the contents of the major JARVIS databases, 4) we demonstrate some of the derived applications, and 5) we discuss outstanding challenges and future work.

**Results and discussion**

**Overview of computational techniques**

There are many computational tools for simulating realistic materials depending on the time and length scales of interest[61]. Before we discuss the details of JARVIS, we will provide a brief list of these techniques and highlight their range of applicability, as summarized in Fig. 1. Relevant techniques include quantum mechanical computations, classical/molecular mechanics, mesoscale modeling, finite element analysis, and engineering design. Each of these methodologies has its own ontology and semantics for describing themselves and the PSPP relationship. For example, 'structure' may imply electronic configurations in the quantum regime, atomic arrangement in molecular mechanics, microstructure, segments in phase field-based mesoscale modeling, and mesh-structure in finite element analysis. Material properties are calculated using corresponding physical laws such as the Schrödinger equation in the quantum regime, or Newton's laws of motion for classical regimes. For realistic material design, it is important to integrate these methods. A major challenge for multiscale modeling is propagating the results of one simulation into another while capturing the relevant physics. Artificial Intelligence (AI) techniques have been applied in each of these domains and can be used to integrate the methods to a certain extent[14]. In JARVIS, we primarily focus on atomistic-based classical and quantum simulations and machine-learning, but we also attempt to integrate other simulation methods with our atomistic data for a few specific applications such as using DFT based elastic constants in orientation distribution function based finite element simulations.

**Software and databases**

The JARVIS infrastructure (Fig. 2) is a combination of databases and tools for running and integrating some of the computational methods mentioned above. The general procedure for adding a dataset to JARVIS is as follows. We start with the goal of finding or designing a material to display or optimize a given property. Then, we decide on an appropriate computational method, as well as a computationally efficient way to screen for the best candidate materials. The screening process can proceed in several steps, with computationally inexpensive methods applied first, followed by more computationally intensive methods on the remaining materials. Whenever possible, the data is compared with available experiments to evaluate the accuracy and quality of the database. Once a large enough dataset is generated, machine learning techniques can be utilized to accelerate the traditional computational approaches.

As an example of making use of multiple computational tools within the same framework, we consider finding materials to maximize solar-cell efficiency. We develop a screening criterion (Spectroscopic Limited Maximum Efficiency, SLME, a part of *JARVIS-Tools*) and calculate the necessary properties (dielectric function and band gap, a part of *JARVIS-DFT*). We test the method by comparing known materials to experiment (*precision and accuracy assessment*), and we perform more accurate meta-GGA and GW calculations (*JARVIS-Beyond DFT*) as additional screening and validation steps. Finally, we develop a machine learning model (*JARVIS-ML*) to accelerate future materials design. Details of this example can be found in Ref. [49,50]. Similar case-studies for thermoelectrics, dielectrics, and infrared-phonon modes are available in Ref. [55] and Ref. [60].

The database component of JARVIS consists of JARVIS-DFT for DFT calculations and JARVIS-FF for molecular dynamics simulations. JARVIS-ML hosts several machine learning models based on our datasets. JARVIS-Tools contains tools for automating, post-processing and disseminating generated data, as well as several derived applications such as JARVIS-Heterostructure. We also include precision and accuracy analyses of the generated data, which consists of comparing DFT data with experiments, comparing FF data with DFT, comparing ML models with DFT, etc. As a lower-level technique (see Fig. 1), JARVIS-DFT data can be fed into JARVIS-FF and JARVIS-ML models, but not vice versa. We use JARVIS-ML to accelerate both JARVIS-DFT and JARVIS-FF. In this way, the JARVIS-infrastructure establishes a joint integration for automation and generation of repositories. We provide several social-media platforms to build a community of interest. Some of the key resources for the JARVIS-infrastructure are shown in Table 1.

**JARVIS-Tools**

JARVIS-Tools is a python-based software package with ≈ 20,000 lines of code and consisting of several python-classes and functions. JARVIS-Tools can be used for a) the automation of simulations and data-generation, b) post-processing and analysis of generated data, and c) the dissemination of data and methods, as shown in Fig. 3. It uses cloud-based continuous integration checking including GitHubAction, CircleCI, TravisCI, CodeCov, and PEP8 linter to maintain consistency in the code and its functionalities. The JARVIS-Tools is distributed through an open GitHub repository: https://github.com/usnistgov/jarvis .

An example python class in JARVIS-Tools is `Atoms`. It uses atomic coordinates, element types and lattice vectors to build an `Atoms` object from which several properties, such as density and chemical formula, can be calculated. This `Atoms` class, along with several other modules (discussed later), can be used for setting up calculations with external software packages. An example of the `Atoms` class is shown in Fig. 4.

The 'Atoms' class along with many other modules in JARVIS-Tools are used to generate input files for automating software codes. Currently, JARVIS-Tools can be used to automate DFT calculations with packages such as Vienna Ab-initio simulation package (VASP)[62,63], Quantum Espresso (QE)[64]; MD with Large-scale Atomic/Molecular Massively Parallel Simulator (LAMMPS)[65]; ML with Scikit-learn[66], Keras[67], and LightGBM[68]; Wannier calculations with Wannier90[69] and Wanniertools[70]. A number of predefined workflows are available in JARVIS-Tools that are continuously being used to calculate properties of uncharacterized or existing materials in the database. Three workflows are shown in Fig. 5. For DFT calculations, an input Atoms class is used to generate input files for VASP (Fig. 5a) with the 'VaspJob' class in order to calculate the desired properties, such as the energy. We automatically perform calculations to converge numerical parameters like the k-points and plane-wave cut-off for individual materials. Geometry optimization is then carried out with energy, force, and stress relaxation. We have chosen a particular set of pseudopotentials or PAWs as tested and recommended by the software developers of various codes. Subsequent properties, such as band structure, dielectric function, elastic constants, piezoelectric constants or spin-orbit spillage are computed on the relaxed structure. Later, custom jobs can also be run on the optimized structure using 'VaspJob', such as Wannier90 calculations using the 'Wannier90Win' class, which generates the input files for an Atom class and a chosen set of pseudopotentials, disentanglement window and other controlling parameters. All of these steps produce a JavaScript Object Notation (JSON) file once the calculations are done as a signature of their completion. The workflows can be restarted from intermediate computations, making the calculations robust to interruptions due to computer failure, etc. We also add several error-handlers in the workflows to automatically re-submit a calculation if a typical error is encountered.

A similar workflow is shown for an example of FF based on LAMMPS calculations in Fig. 5b. Here, for a particular force-field such as Ni-Al[58], for example, all the structures related to Ni, Al and Ni-Al are obtained from the DFT database and converted into a LAMMPS input format using '`Atoms`', '`LammpsData`' and '`LammpsJob`' objects. Then a series of geometry optimization, vacancy formation energy, surface energy, and phonon-related calculations are run, based on the symmetry of the structure. All of these steps use a set of ".mod" module files with input parameters that control respective LAMMPS calculations. The obtained results are compared with corresponding DFT data, to evaluate the quality of an FF for a particular system or simulation.

In machine learning calculations, the input materials-data is transformed into several machine-readable descriptors[71] such as `CFID` dataset or STM image 'numpy' arrays. As we are not going to generate another set of data for testing ML models, we split the dataset into training and testing sets in a 90: 10 or similar split. Using k-fold cross-validation, we obtain hyperparameters for the chosen algorithm, for example, the number of trees, learning rate, etc. in the case of Gradient Boosting Decision Tree (GBDT). We choose the optimized parameters and train on 90 % train data and test on the 10 % test data to evaluate the truly predictive performance on unseen data. We also carry out k-fold cross-validation using the finalized model to get model uncertainty. Later, we can analyze interpretability with techniques such as feature importance in tree-based algorithms or filters in neural networks. These models are saved in Pickle, cPickle and Joblib modules for model persistency. We also carry out uncertainty analysis using methods such as prediction interval and Monte-Carlo dropouts[72]. A few examples and Jupyter notebooks are provided on the GitHub page to illustrate the above-mentioned methods. More details about the individual python modules mentioned above can be found in the JARVIS-Tools documentation (https://jarvis-

tools.readthedocs.io/en/latest/). A documentation on integrating JARVIS-Tools with the database is available at (https://jarvis-materials-design.github.io/dbdocs/).

After running the automated calculations, the data is post-processed to predict various material properties (such as bandgap, formation energy, spin-orbit spillage, SLME, density of states, phonons, dielectric function, or STM image). Many of the python classes use '`ToDict`' and '`FromDict`' methods that help store the metadata. These metadata are then used with HTML[73], Javascript, Flask[74] and other related software to make web-pages and web-apps. The metadata is also shared in public repositories such as Figshare (https://figshare.com/authors/Kamal_Choudhary/4445539 ), and JARVIS-Representational state transfer (REST) API, based on the MGI philosophy of creating and using interoperable datasets. Note that through the JARVIS-REST API, a user can download JARVIS data and can also upload/store their own data. If the stored data follows the schema (in XSD format), then the API automatically generates HTML pages for the user's data. The data generated in JARVIS is mainly stored in Extensible Markup Language (XML), JavaScript Object Notation (JSON), Comma-Separated Values (CSV) or American Standard Code for Information Interchange (ASCII) format and, again, JARVIS-Tools can be used to analyze the pre-calculated data for materials design. A wrapper-code for the REST-API upload and download is available at (https://github.com/usnistgov/jarvis/blob/master/jarvis/db/restapi.py). An example of downloading precalculated dataset with JARVIS-Tools is shown in Fig. 4. JARVIS-Tools, along with the various software shown in Fig. 3, has led to several databases shown in Fig. 6.

**JARVIS-DFT**

Density functional theory is one of the most commonly used techniques in condensed-matter physics to solve real-world materials problems. In DFT, instead of solving the fully interacting

Schrödinger equation, we solve the Kohn-Sham equations, which describe an effective non-interacting problem, greatly improving computational efficiency. Although exact in principle, DFT requires several approximations in practice. In particular, various levels of approximation to the exchange-correlation functional are possible, which require different computational effort. Most existing DFT databases use the common GGA-PBE throughout all the material-classes. JARVIS-DFT can be viewed as an attempt to build a repository beyond existing DFT databases. JARVIS-DFT[7,8,25-27,30,31,49,53,55] was started in 2017 and contains data for ≈ 40,000 materials, with ≈ 1 million calculated properties, mainly based on the VASP package. Although there are several DFT-functionals adopted in JARVIS-DFT, we use vdW-DF-OptB88 consistently for all the 3D, 2D, 1D and 0D materials. This functional has been shown to provide accurate predictions for lattice-parameters and energetics for both vdW and non-vdW bonded materials[30]. In addition to hosting 3D bulk materials, the database consists of 2D monolayer, 1D-nanowire, and 0D-molecular materials (as shown in Table 2). However, to date, 3D and 2D materials have primarily been distributed publicly. Moreover, other exchange-correlation functionals are considered (as shown in Table 3), which can help estimate the prediction uncertainty. While vdW-DF-OptB88 can predict accurate lattice parameters and formation energies, bandgaps are still underestimated. Calculations with hybrid functionals (such as range-separated HSE06 and PBE0) and many-body approaches (such as $G_0W_0$) remain too computationally expensive[23] to use in a high-throughput methodology for thousands of materials. Hence, a meta-GGA Tran-Blaha-modified Becke-Johnson (TBmBJ) potential is used to provide a good balance between computational expense and accuracy. The TBmBJ accuracy is shown to be close enough to the high-level methods such as HSE06 at up to ten times lower computational expense[57]. Accurate prediction of optical gaps by calculation of the frequency-dependent dielectric function is important for several applications, for example, solar-cell efficiency calculations. Accurate prediction of bandgaps also helps in obtaining accurate frequency-dependent dielectric functions, which can be critical for solar-cell efficiency calculations; however, TBmBJ cannot describe the excitonic nature of electron-hole pairs in low-dimensional materials. In addition to TBmBJ, we are generating HSE06, PBE0, $G_0W_0$ and DMFT datasets, which can be considered as beyond-DFT methods discussed in the next section. Next, SOC is varied to analyze the differences introduced by this coupling. These differences are used to discover 3D and 2D topological materials. In addition, several DFT databases are developed including properties such as frequency-dependent dielectric function and electric field gradient. A few important protocols such as k-point automatic convergence are also introduced. A snapshot of the JARVIS-DFT website along with a list of properties that are available is shown in Fig. 7. JARVIS-DFT has several filtering options on the website to screen candidate materials. We provide the input files as downloadable .zip files, especially for the users who do not have much expertise in using python-based codes. Raw input and output files (on the order of 1 terabyte) will soon be made publicly available through the Figshare repository, NIST-Materials data repository, and Materials Data Facility (MDF). A summary table, with the number of data available with vdW-DF-OptB88 and other methods, is shown in Tables 2 through 4. Table 2, Table 3 and Table 4 provides a summary of available materials classes, DFT functionals used and materials properties available in the JARVIS-DFT database.

**JARVIS-Beyond-DFT**

While quantum mechanical methods in single-particle theories such as DFT or DFT+U methods (mainly GGA) are fast and can predict accurate results for most structural parameters, even when relatively strong electron correlations are present, qualitative predictions of excited state properties may require beyond-DFT methods[75]. Beyond-DFT calculations have been applied to many materials systems, including cuprates and Fe-based high-temperature superconductors, Mott insulators, heavy Fermion systems, semiconductors, photovoltaics, and topological Mott insulators[75]. In the last few decades, both perturbative and stochastic approaches have been developed to understand these strongly correlated materials. These approaches, including Dynamical Mean Field Theory (DMFT)[76], the GW approximation, , or hybrid exchange-correlation functionals are often called beyond-DFT methods since they go beyond the limit of semilocal DFT. The materials design community often requires benchmarking for particular cases, where it is necessary to use beyond-DFT methods, in order to assess accuracy of the results. In the JARVIS-Beyond-DFT database we are building a database of spectral functions and related quantities as computed using meta-GGA, GW, hybrid functionals, and LDA+DMFT for head-to-head comparison on 100+ materials.

In the JARVIS-Beyond-DFT[75] database we try to answer a few key questions regarding discoveries through a materials database for quantum materials. First, where is it necessary to use a beyond-DFT method, and which method to be use? Second, how do different "beyond-DFT" methods compare with experiments? Target materials include but are not limited to various transition metal oxides, perovskites and mixed perovskites, nickelates, transition metal dichalcogenides, and a wide range of metals starting from alkali metals to transition metals, and various Iron-based superconductors. JARVIS-Beyond-DFT will be distributed through the website: https://jarvis.nist.gov/jarvisbdft/.

**JARVIS-FF**

Classical force-field-/interatomic-potential-based simulations are the workhorse technique for large scale atomistic simulations. They are especially suited for temperature-dependent and defect-related phenomena. Several varieties of FFs differ based on the materials system and the underlying phenomena under investigation, e.g., whether they include bond-angle information and fixed or dynamic charges. Also, they are generally designed for particular applications and phases, making it difficult to ascertain whether they will perform well in simulations for which they were not explicitly trained. JARVIS-FF[27,58] is a collection of LAMMPS calculation-based data consisting of crystal structures, formation energies, phonon densities of states, band structures, surface energies and defect formation energies. There are ≈ 110 FFs in the database, for which the corresponding crystal structures are obtained from JARVIS-DFT, converted to LAMMPS format inputs, and used in a series of LAMMPS calculations to produce the aforementioned properties. These properties, when compared with corresponding DFT data, can help a user analyze the quality of a force-field for a particular application. Examples include the comparison of DFT convex hull with FF, elastic modulus, surface energy and vacancy formation energy data. Some types of FFs included are EAM, MEAM, Bond-order and Tersoff, COMB, and ReaxFF as shown in Table. 5. Furthermore, we plan to include several recently developed machine learning force-fields into JARVIS-FF. A snapshot of the JARVIS-FF website is also shown in Fig. 8.

**JARVIS-ML**

Machine learning has several applications in materials science and engineering[14,80,81], such as automating experimental data analysis, discovering functional materials, optimizing known ones by accelerating conventional methods such as DFT, automating literature searches, discovering physical equations, and efficient clustering of materials and their properties. There are several data types that can be used in ML such as scalar data (e.g., formation energies, bandgaps), vector/spectra data (e.g., density of states, dielectric function, charge density, X-ray diffraction patterns, etc.), image-based data (such as scanning tunneling microscopy and transmission electron microscopy images), and natural language processing-based data (such as scientific papers). In addition, ML can be applied on a variety of materials classes such as bulk crystals, molecules, proteins and free-surfaces.

Currently, there are two types of data that are machine-learned in JARVIS-ML[49,53,55,59,60]: discrete and image-based. The discrete target is obtained from the JARVIS-DFT database for 3D and 2D materials. There have been several descriptor developments as attempts to capture the complex chemical-structural information of a material[71]. We compute CFID descriptors for most crystal structures in various databases (as shown in Table. 6). Many of these structures are non-unique but can still be used for pre-screening applications[49]. The CFID can also be applied to other materials classes such as molecules, proteins, point defects, free surfaces, and heterostructures, which are currently ongoing projects. These descriptor datasets, along with JARVIS-DFT and other databases, act as input and outputs for machine learning algorithms. The CFID consists of 1557 descriptors for each material: 438 average chemical, 4 simulation-box-size, 378 radial charge-distribution, 100 radial distribution, 179 angle-distribution up to first neighbor, and another 179 for the second neighbor, 179 dihedral angle up to fist neighbor and 100 nearest neighbor descriptors. More details can be found in Ref. [59]. Currently, we provide CFID descriptors only, but other descriptors such as Coulomb-matrix, and sine-matrix will be provided soon. With CFID descriptors, we trained several classification and regression tasks. Once these models are trained, parameters are stored that can predict the properties of an arbitrary compound quickly. We developed a web-based application to host the trained models, as shown in Fig. 9, and a list of the trained properties are displayed there as well. We note that classical quantities such as bulk modulus, maximum infrared (IR) active mode, and formation energies can be accurately trained, especially with regression models. For other properties such as bandgaps, magnetic moments, piezoelectric coefficients, thermoelectric coefficients, high accuracy models are obtained for classification tasks only. In addition to the descriptor-based data, we develop Scanning Tunneling Microscopy (STM)[53] image classification models that can be used to accelerate the analysis of STM data. The images are converted into a black/white image to identify spots with/without atoms. The model's accuracy is compared with respect to DFT data or experiments wherever applicable.

**Derived apps**

The knowledge developed through the above-mentioned databases and tools can serve as static content, as well as accessed through dynamic user-defined inputs. Derived applications (apps) are designed to help a user analyze the combinatorics in the data. Based on the databases and tools discussed above, several apps are derived from JARVIS such as JARVIS-Heterostructure[31], JARVIS-Wannier TB, and JARVIS-ODF. JARVIS-Heterostructure (as shown in Fig. 10a) can be used to characterize heterojunction type and modeling interfaces for exfoliable 2D materials. We classify these heterostructures into type-I, II and III systems according to Anderson's rule, which is based on the band-alignment with respect to the vacuum potential of non-interacting monolayers, obtained from JARVIS-DFT. The app also generates crystallographic positions for the heterostructure that could be used as input for subsequent calculations. JARVIS-WannierTB (as shown in Fig. 10b) can be used to solve Wannier Tight Binding Hamiltonians on arbitrary k-points for 3D and 2D materials. Properties such as the band structure and the density of states can be predicted on the fly from this app. Additionally, many other apps are being developed, which are primarily based on the Flask python package[74].

The JARVIS-ODF (Orientation Distribution Function) library is under development, which aims to calculate volume-averaged (meso-level) material properties, including the elasto-plastic deformation behavior, using the property data available for single crystals in the JARVIS database.

Once generated, the JARVIS-ODF library will be capable of obtaining such material properties for all crystalline structures.

**Accuracy and precision analysis**

In simulations, accuracy refers to the degree of closeness between a calculated value and a reference value, which can be from an experiment or a high-fidelity theory. Precision refers to the degree of closeness between numerical approaches to solving a certain model, including the effect of convergence and other simulation parameters.

In JARVIS-DFT, the accuracy of the DFT data is obtained by comparing it to available experimental results (see Supplementary Table 1-9). The accuracy of JARVIS-FF and JARVIS-ML, instead, is given with respect to DFT results. Note that the numbers of high-quality experimental measurements or high-fidelity calculations for a given property are often low. Therefore, the accuracy metrics we derive in our works are obtained only for the few cases we can directly compare, not for the entire dataset. In Table. 7, we provide accuracy metrics for some material properties in the JARVIS-DFT with respect to experiments. In addition to the scalar data, vector/continuous data, such as frequency dependent dielectric function and Scanning Tunneling Microscopy (STM) images, are compared to a handful of experimental data points as well. Details of individual properties can be found in Ref.[8,30,49,52,53,55,59,60]

JARVIS-FF data accuracy is calculated with respect to the DFT data, for properties such as the convex hull, bulk modulus, phonon frequencies, vacancy formation energies and surface energies. In Refs.27,58, we showed this through several examples, including the comparison of Ni-Al and Cu-O-H systems convex hulls to DFT data. We also showed examples of comparing defect formation energies, surface energies and its effects on Wulff-shape. Although these accuracy analyses are based on 0K DFT data, they are useful in predicting temperature-dependent and dynamical behavior because we consider several crystal prototypes of a system.

JARVIS-ML model accuracy is evaluated on the test-set (usually 10 %) representing previously unseen DFT data for both regression and classifications models. Accuracy of regression and classification models are reported in terms of mean absolute error (MAE) and Receiver Operating Characteristic (ROC) Area Under Curve (AUC) metric respectively. A brief summary of regression and classification model accuracy results is given below in Table. 8 and 9. Details of the accuracy analyses are provided in Refs. [49,53,55,59,60]

Precision analysis can refer to a wide variety of optional selections of simulation set-ups. Examples of precision analysis in JARVIS-DFT are using our convergence protocols for k-points and plane-wave cutoff, and the convergence of Wannier tight-binding Hamiltonians. Using a converged k-point mesh and plane-wave cutoff[28] for each individual material is necessary to obtain high-quality data. Note that these DFT convergences are carried out for energies of the system only, and not for other properties. However, we impose tight convergence parameters for both k-points and energy cutoff (0.001 eV/cell), which typically results in other physical quantities being converged as well. In JARVIS-FF, comparison across structure-minimization methods for calculating surface and vacancy formation energy values are examples of precision analysis[27]. We find that the FF simulation setups ('refine' and 'box' methods) have minimal effect on the FF-based predictions. For classification ML models, precision is the ratio $\frac{TP}{TP + FP}$ where TP is the number of true positives and FP the number of false positives, which can be derived from the confusion. Precision analysis for classification ML model for STM Bravais-lattices are available in Ref. [53]. We find high precision (more than 0.87) for all of the 2D-Bravais lattices. Precision analysis for regression tasks are still ongoing and will be available soon.

**Future work**

Given that the number of all possible materials [77] could be of the order of $10^{100}$, and furthermore existing materials properties can be computed at increasing levels of accuracy/cost, the JARVIS databases will always be incomplete. This represents an opportunity for JARVIS to be drastically expanded in the future. Future work will be aimed at addressing some of the limitations of the existing databases, and may include additions like defect/disorder properties, magnetic ordering, non-linear optoelectronics, more beyond-DFT calculations, temperature-dependent properties, integration with experiments, and more detailed uncertainty analysis. Moreover, several ML models and methods for data-prediction and uncertainty quantification will be developed for 'explainable AI' (XAI) and transfer-learning (TL)-based research. Other derived apps such as JARVIS-ODF, JARVIS-Beyond-DFT, JARVIS-GraphConv, and JARVIS-STM are also being developed. In addition to the technical aspects, the broader impact of the infrastructure will be to provide a research platform that will allow maximum participation of worldwide researchers. NIST-JARVIS currently hosts pre-computed data and would host on-the-fly calculation resources also. To make the data-processing user-friendly, we have a few filtering options on the JARVIS-DFT website. Furthermore, advanced filtering tools will be available through ElasticSearch package soon. ElasticSearch integration will allow cross-filtering among several databases. We are also working on several visualization tool integration using Plotly, Javascript and XSLT which will be available on the web soon.

In summary, we described the Joint Automated Repository for Various Integrated Simulations (JARVIS) platform, which consists of several databases and computational tools to help accelerate materials design and enhance industrial growth. JARVIS includes three major databases: JARVIS-DFT for density functional theory calculations, JARVIS-FF for classical force-field calculations, and JARVIS-ML for ML predictions. In addition, we provide JARVIS-Tools, which is used to generate the databases. The generated data is provided publicly with several example notebooks, documentation and calculation examples to illustrate different components of the infrastructure. We believe the publicly available data and resources provided here will significantly accelerate futuristic materials-design in various areas of science and technology.

**Methods**

The entire study was managed, monitored, and analyzed using the modular workflow, which we have made available[54] on our JARVIS-Tools GitHub page (https://github.com/usnistgov/jarvis).

**Density functional theory calculations**

The DFT calculations are mainly carried out using the Vienna Ab-initio simulation package (VASP)[62,63]. We use the projected augmented wave method and OptB88vdW functional[56], which gives accurate lattice parameters for both van der Waals (vdW) and non-vdW solids[30]. Both the internal atomic positions and the lattice constants are allowed to relax in spin-unrestricted calculations until the maximal residual Hellmann–Feynman forces on atoms are smaller than 0.001 eV Å$^{-1}$ and energy-tolerance of $10^{-7}$ eV. We do not consider magnetic orderings besides ferromagnetic yet, because of a high computational cost. We note that nuclear spins are not explicitly considered during the DFT calculations. The list of pseudopotentials used in this work is given on the GitHub page. The k-point mesh and plane-wave cut-off were converged for each material using the automated procedure described in Ref[28]. The elastic constants are calculated using the finite difference method with six finite symmetrically distinct distortions. The thermoelectric coefficients such as power factor and Seebeck coefficients are obtained with the BoltzTrap code with Constant Relaxation Time approximation (CRTA)[78]. Optoelectronic properties such as dielectric function and solar-cell efficiency are calculated using linear-optics methods mainly using OptB88vdW and TBmBJ. We also compared such data with HSE06 and $G_0W_0$. The piezoelectric, dielectric and phonon modes at Γ-point are calculated using Density Functional Perturbation Theory (DFPT). Topological spillage for identifying topologically non-trivial materials is calculated by comparing DFT wave functions with/without SOC[7,26]. 2D exfoliation energies are calculated by comparing bulk and 2D monolayer energy per atom. The 2D heterostructure[31] behavior is predicted using Zur and Anderson methods. Wannier tight binding Hamiltonians are generated using the Wannier90 code[69]. 2D STM images are predicted using the Tersoff-Hamman method[53].

**Force-field calculations**

Classical force-field calculations are carried out with the LAMMPS software package[65]. In our structure minimization calculations, we used $10^{-10}$ eVÅ$^{-1}$ for force convergence and 10000 maximum iterations. The geometric structure is minimized by expanding and contracting the simulation box with 'fix box/relax' command and adjusting atoms until they reach the force convergence criterion. These are commonly used computational set-up parameters. After structure optimization point vacancy defects are created using Wycoff-position data. Free surfaces for maximum miller indices up to 3 are generated. The defect structures were required to be at least 1.5 nm long in the *x*, *y* and *z* directions to avoid spurious self-interactions with the periodic images of the simulation cell. We enforce the surfaces to be at least 2.5 nm thick and with 2.5 nm vacuum in the simulation box. The 2.5 nm vacuum is used to ensure no self-interaction between slabs, and the slab-thickness is used to mimic an experimental surface of a bulk crystal. Using the energies of perfect bulk and surface structures, surface energies for a specific plane are calculated. We should point out that only unreconstructed surfaces without any surface-segregation effects are computed, as our high-throughput approach does not allow for taking into account specific, element dependent reconstructions yet. Phonon structures are generated mainly using the Phonopy package interface[79].

**Machine learning training**

Machine learning models are mainly trained using Scikit-learn[66], Keras[67], and LightGBM[68] (TensorFlow backend) software. For DFT generated scalar data such as formation energies, bandgaps, exfoliation energies etc. the crystal structures are converted into a Classical Force-field Inspired Descriptors (CFID) input array and the DFT data is used as target data, which is then train-test split in a ratio of 90: 10. Preprocessing such as 'VarianceThreshold', 'StandardScalar' are used before ML training. Regression models' performance are generally reported in terms of Mean Absolute Error (MAE) or $r^2$, while that for classification models using the Receiver Operating Characteristic (ROC) Area Under Curve (AUC) value which lie between 0.5 and 1.0. Several other analyses such as feature importance, k-fold cross validation and learning curve are carried out after the model training. The trained model is saved in pickle and joblib formats for model persistence. All the web-apps are developed using JavaScript, Flask and Django packages[74].

**Data availability**

JARVIS-related data is available at the JARVIS-API (http://jarvis.nist.gov), JARVIS-DFT (https://jarvis.nist.gov/jarvisdft/ ), JARVIS-FF (https://jarvis.nist.gov/jarvisff/ ), JARVIS-ML (https://jarvis.nist.gov/jarvisml/ )  websites. The metadata is also available at the Figshare repository, see https://figshare.com/authors/Kamal_Choudhary/4445539 .

**Code Availability**

Python-language based codes with examples are available at JARVIS-Tools page: https://github.com/usnistgov/jarvis .

**Acknowledgements**

K.C., K.F.G., and F.T. thank the National Institute of Standards and Technology for funding, computational, and data-management resources. K.C. thanks the computational support from XSEDE computational resources under allocation number TG-DMR 190095. Contributions from KC were supported by the financial assistance award 70NANB19H117 from the U.S. Department of Commerce, National Institute of Standards and Technology. Contributions by S.M., K.H., K.R., and D.V. were supported by NSF DMREF Grant No. DMR-1629059 and No. DMR-1629346. X.Q. was supported by NSF Grant No. OAC-1835690. B.G.S and S.V.K acknowledge work performed at the Center for Nanophase Materials Sciences, a US Department of Energy Office of Science User Facility. A.A. acknowledges partial support by CHiMaD (NIST award # 70NANB19H005). G.P. was supported by the Los Alamos National Laboratory's Laboratory Directed Research and Development (LDRD) program's Directed Research (DR) project #20200104DR. K.C. thanks for helpful discussion with several researchers including Faical Y. Congo, Daniel Wheeler, James Warren, Carelyn Campbell, Chandler Becker, Marcus Newrock, Ursula Kattner, Kevin Brady, Lucas Hale, Eric Cockayne, Philippe Dessauw from National Institute of Standards and Technology; Karen Sauer, Igor Mazin, Nirmal Ghimire, Patrick Vora from George Mason University; Rama Vasudevan, Maxim Ziatdinov from Oak Ridge National Lab, Deyu Lu and Matthew Carbone from Brookhaven National Lab; Marnik Bercx, Dirk Lamoen from University of Antwerp; Yifei Mo from University of Maryland; Anubhav Jain and Sinead Griffin from Lawrence Berkeley National Laboratory; Surya Kalidindi from Georgia Tech.; Tyrel McQueen and David Elbert from Johns Hopkins University; Richard Hennig from University of Florida; Giulia Galli and Ben Blaiszik from University of Chicago; Qiang Zhu from University of Nevada-Las Vegas; Dilpuneet Aidhy from University of Wyoming; Susan B. Sinnott, Tao Liang from Pennsylvania State University.

**Author Contributions**

KC designed the JARVIS workflows, carried out high-throughput calculations, analysis and developed the websites. FT contributed to the development of k-point and other convergence protocol, Beyond-DFT development and several other analyses. KG contributed to the development of topological materials discovery and Wannier-tight binding Hamiltonian projects. ACER assisted in the deployment of the web-apps. BDC, AA and AGK contributed to the machine-learning tasks. AJB, AHR, AC, VS, AD contributed to the phonon data analysis. ZT contributed to the development of the JARVIS-API website. JHS contributed to the experimental validation of some of the screened materials. JJ and RP contributed in the solar-cell and topological materials discovery tasks. GC, ER, XQ, HZ, SVK, BS, GP contributed to the discovery and characterization of low-dimensional materials. PA contributed to the elastic constant analysis task. SM, KR, DV and KH contributed to the Beyond-DFT project. All authors contributed to writing the manuscript.

**Competing interests**

The authors declare no competing interests.

**Figure captions**

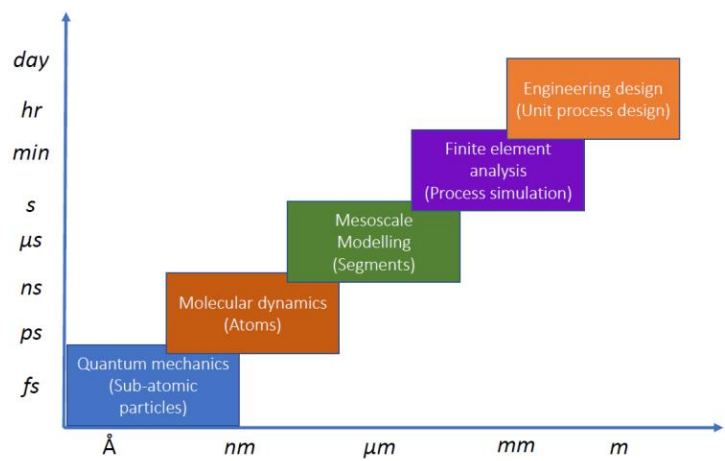

Fig. 1 **Length and time-scale based computational materials design techniques.** We primarily focus on the lowest two levels of the computational methodologies, DFT and MD, but we integrate with other simulation methods for specific applications.

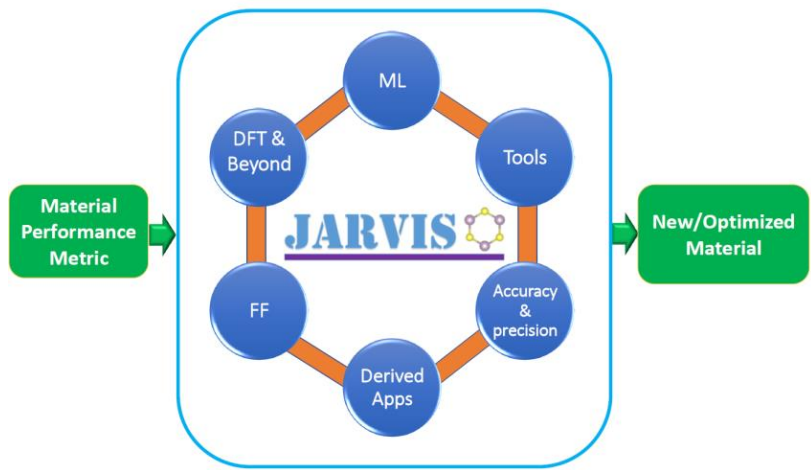

Fig. 2 **An overview of the JARVIS infrastructure.** For a given materials performance metric, several JARVIS components can work together to design optimized or completely new materials.

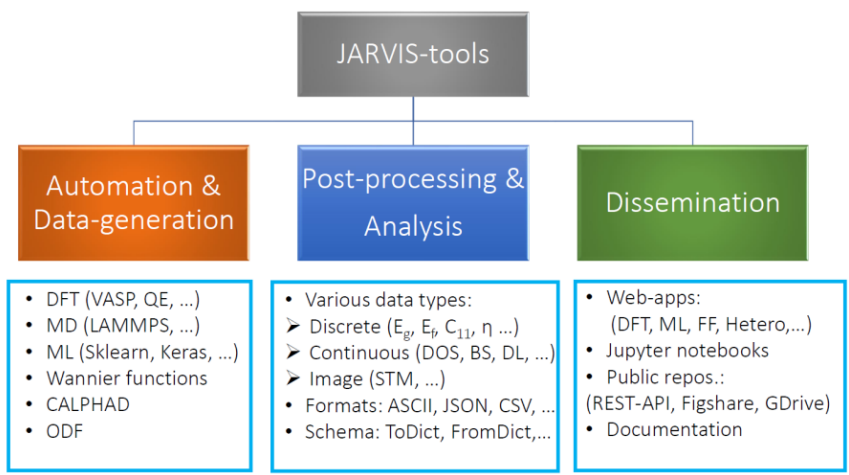

Fig. 3 **Three main components of the JARVIS-Tools package and their capabilities**.

```
>>> from jarvis.core.atoms import Atoms
>>> box = [[2.715, 2.715, 0], [0, 2.715, 2.715], [2.715, 0, 2.715]]
>>> coords = [[0, 0, 0], [0.25, 0.25, 0.25]]
>>> elements = ["Si", "Si"]
>>> Si = Atoms(lattice_mat=box, coords=coords, elements=elements)
>>> density = round(Si.density,2)
>>> print (density)
2.33
>>>
>>> from jarvis.db.figshare import data
>>> dft_3d = data(dataset='dft_3d')
>>> print (len(dft_3d))
36099
>>> from jarvis.io.vasp.inputs import Poscar
>>> for i in dft_3d:
...     atoms = Atoms.from_dict(i['atoms'])
...     poscar = Poscar(atoms)
...     jid = i['jid']
...     filename = 'POSCAR-'+jid+'.vasp'
...     poscar.write_file(filename)
>>> dft_2d = data(dataset='dft_2d')
>>> print (len(dft_2d))
1070
>>> for i in dft_2d:
...     atoms = Atoms.from_dict(i['atoms'])
...     poscar = Poscar(atoms)
...     jid = i['jid']
...     filename = 'POSCAR-'+jid+'.vasp'
...     poscar.write_file(filename)
```

Fig. 4 **Examples of using python classes in JARVIS-Tools for constructing 'Atoms' class and downloading data.** More tutorial-based examples are available on the documentation pages.

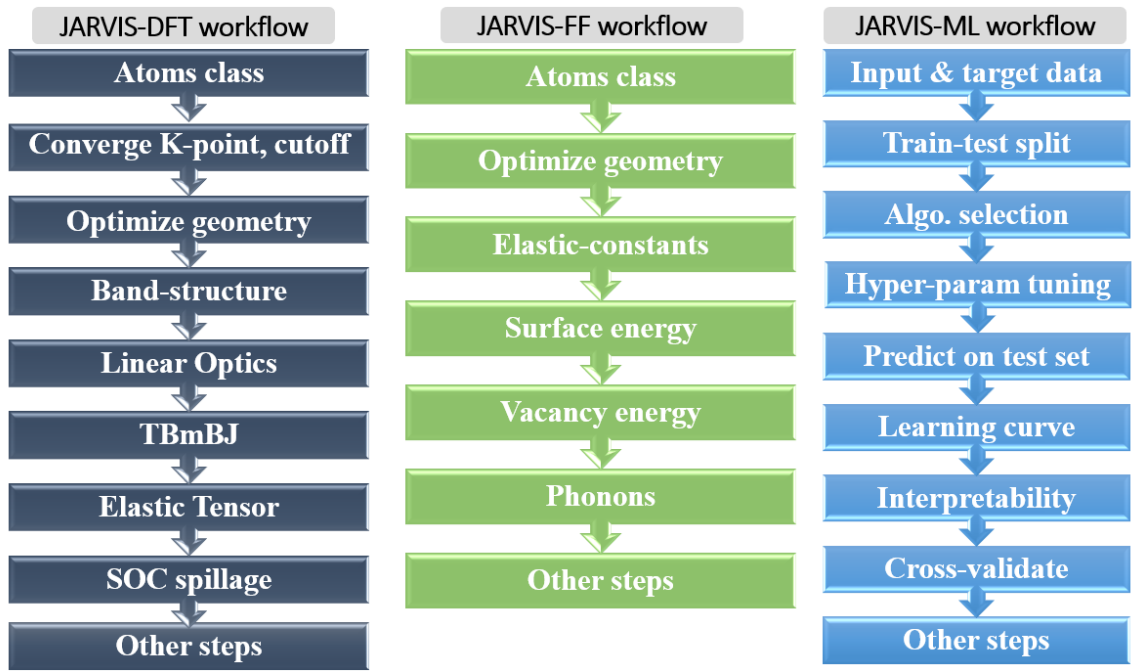

Fig. 5 **Flowcharts showing some of the main steps used in most-commons calculations.** a) JARVIS-DFT, b) JARVIS-FF and c) JARVIS-ML workflows.

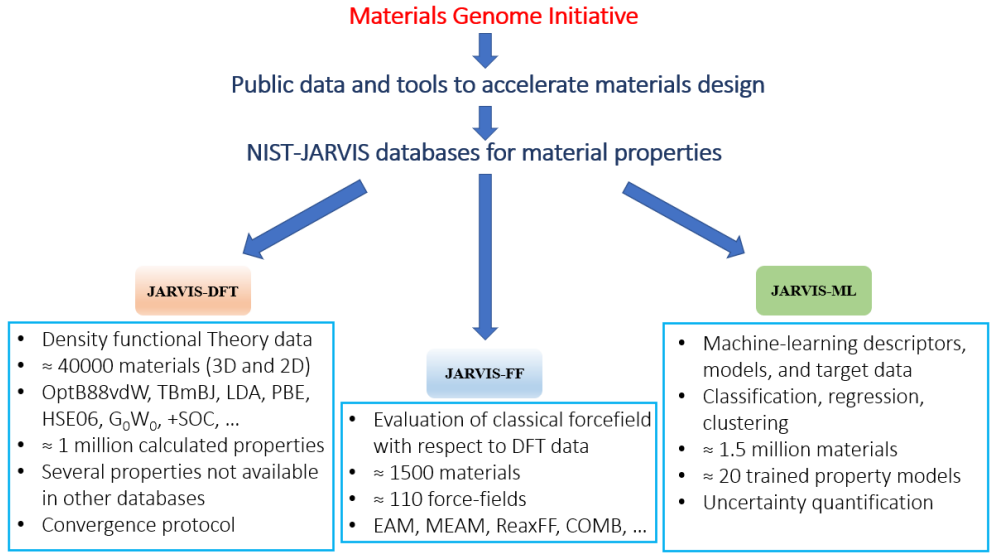

Fig. 6 **Three main databases in JARVIS and a summary of their contents**.

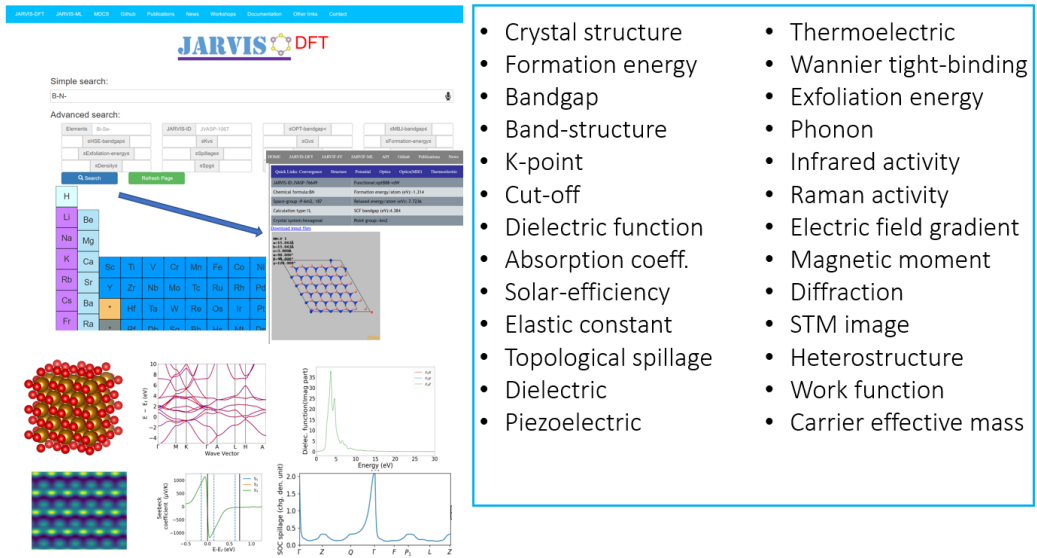

Fig. 7 **A snapshot of JARVIS-DFT website and summary of its contents**.

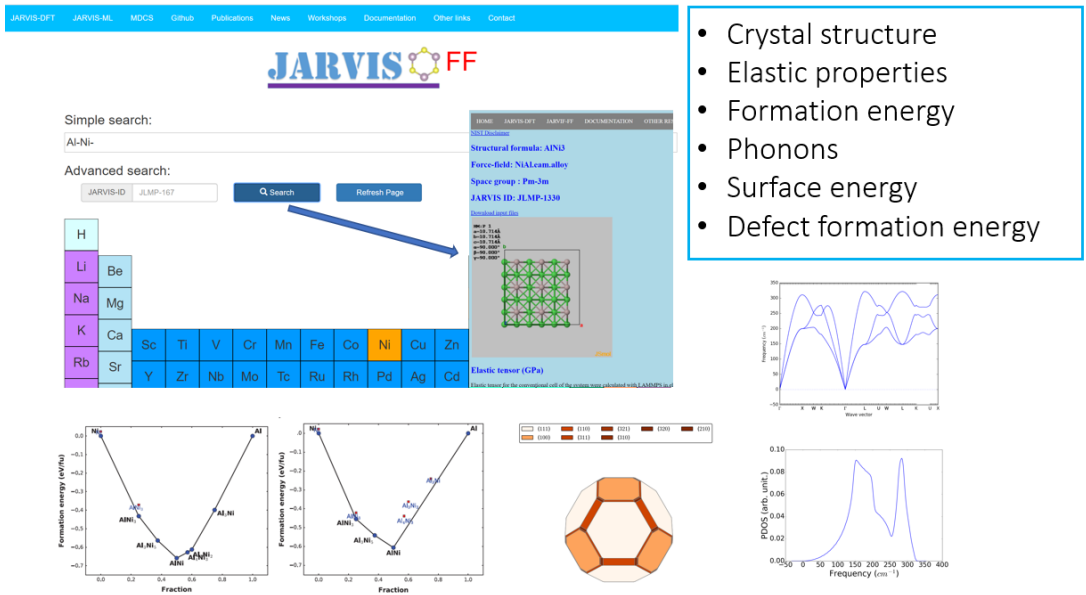

Fig. 8 **A snapshot of JARVIS-FF website and summary of its contents**.

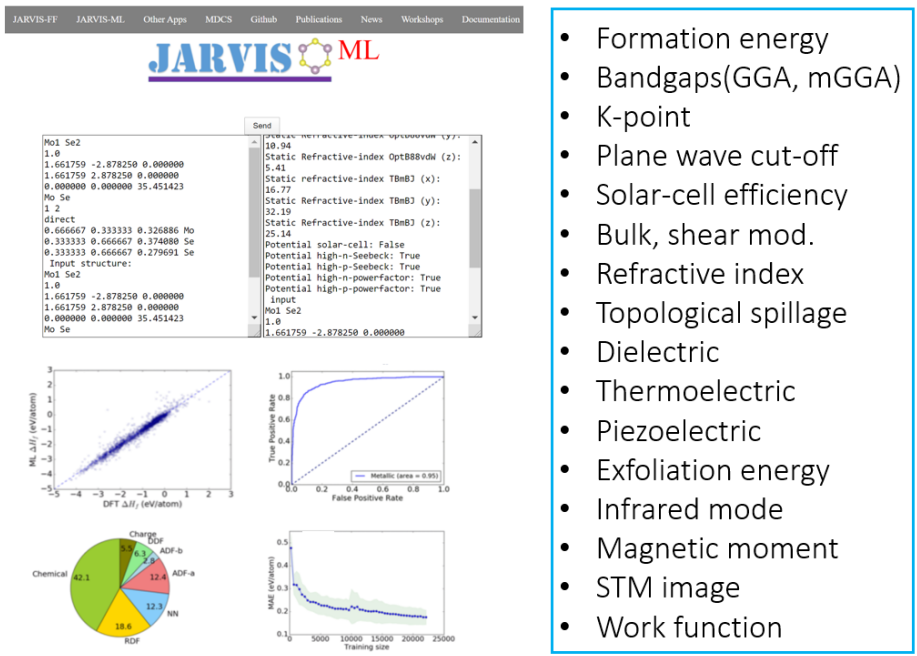

Fig. 9 **A snapshot of JARVIS-ML website and summary of its contents**.

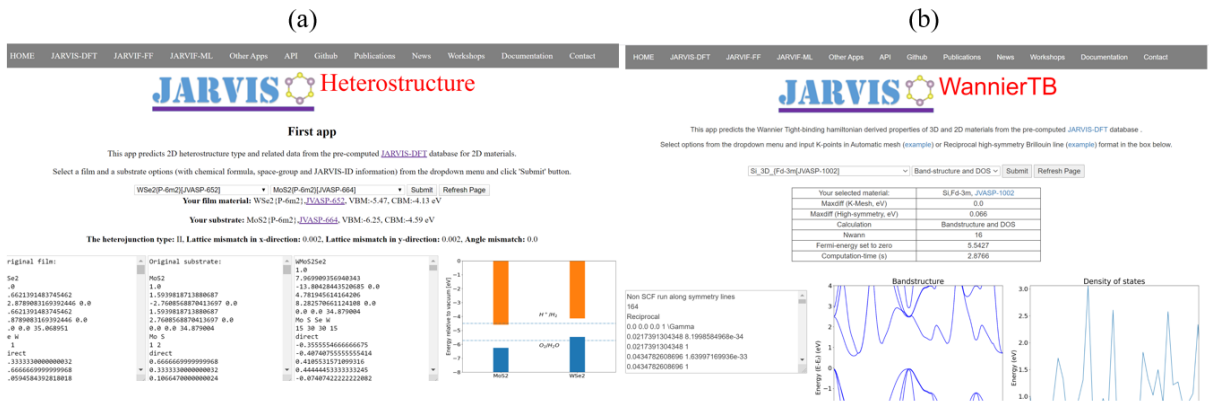

Fig. 10 **Snapshots of JARVIS-DFT derived apps**. a) JARVIS-Heterostructure and b) JARVIS-Wannier Tight Binding.